# Polarization-decoupled Flat Displays

Isma Javed[1], Muhammad Ashar Naveed[1], Muhammad Qasim Mehmood[1], Yehia Massoud[2]

[1]MicroNano Lab, Electrical Engineering Department, Information Technology University (ITU) of the Punjab, Ferozepur Road, Lahore 54600, Pakistan.
[2]Innovative Technologies Laboratories (ITL), King Abdullah University of Science and Technology (KAUST), Saudi Arabia.

*Abstract*—**Many modern applications like entertainment displays, data encryption, security, and virtual reality (VR) technology require asymmetric light manipulation. Symmetric spin-orbit interactions (SOI) apply a limit in achieving an asymmetrical metahologram. However, different reported asymmetric SOI's based on propagation and geometric phase mergence techniques effectively break this limit at the expense of design complexity and greater computation cost. This work proposes a novel helicity multiplexing technique that breaks all the aforementioned barriers in achieving on-axis dual side holograms. Benefiting from the geometric phase modulation of anisotropic nano-resonating antennas, we have employed a single unit cell to achieve helicity multiplexing. A low extinction coefficient material a-Si:H is used for device analysis. Due to the simple single unit cell-based designing technique, simulation and fabrication complexities were significantly reduced. As a result, based on the helicity and incidence direction of electromagnetic wave, we have achieved highly transmissive dual holographic images in the visible band. Our simulated efficiencies are 55%, 75%, and 80% for the blue (λ = 488 nm), green (λ = 532 nm), and red light (λ = 633 nm).**

## I. INTRODUCTION

The 2D ultra-thin version of the metamaterial, metasurfaces bring new innovation in the world of applied electromagnetic and photonics by catching unprecedented control over light properties (i.e., amplitude, phase, and polarization) [1, 2]. They have demonstrated numerous interesting applications of dual side displays [3], data encryption [4, 5], beam engineering [6], and wave absorbing [7, 8], etc. Metasurfaces as a holographic optical element (HOE) congregated a prestigious amount of attention due to ultra-compactness, large viewing angle, and high diffraction efficiency. Primarily, for both reflection and transmission mode, plasmonic meta holograms are demonstrated, including multi-color metaholograms, broadband multi-plane holograms, and multiplexed holograms [9, 10]. However, their applicability is greatly constraint by inherent ohmic losses due to metallic resonators, leading to lower efficiency and fabrication defects at optical wavelengths [11].

To overcome these efficiency barriers, high refractive index dielectrics with low extinction coefficients like silicon (Si), titanium dioxides ($TiO_2$), and gallium nitride (GaN) become the more suitable choice. These dielectrics have multiple electric and magnetic resonances due to their inherent properties supporting high transmission in optical regime [12]. Huang *et al.* utilized Si-based nano-rectangular bars for a transmission type all-dielectric metasurface to realize broadband hologram for an optical regime with a total efficiency of 3% [13]. Zhou *et al.* improved the efficiency up to 47% at a single optical wavelength, i.e., 532 nm using crystalline Si and achieved complete 0 to $2\pi$ phase control [14]. Though Si provided fabrication ease and showed promising results compared to plasmonic materials, its performance is still degraded due to the high absorption coefficient at optical wavelengths. As a result, researchers exploited GaN and $TiO_2$ that provide a wide range of transparent windows for optical wavelengths [15, 16]. For example, Chen *et al.* realized a GaN-based multiplexed color router with metalens for three optical wavelengths of 488 nm, 532 nm, and 633 nm with efficiencies of 87%, 91.6% 50.6%, respectively [17]. Chen *et al.* exploited $TiO_2$ based meta-axicons for multiple order bessel beam generation for visible regime [18]. Although these dielectric materials demonstrate promising results, high aspect ratios and incompatibility with existing CMOS technology lead to fabrication challenges limiting their practical applicability [19]. To outplay aforementioned drawbacks, the hydrogen added amorphous silicon (a-Si:H) is introduced, which is well compatible with already mature CMOS foundry. Ansari *et al.* demonstrate the helicity multiplexed hologram with maximum efficiency of 74% at optical wavelength 633nm [20].

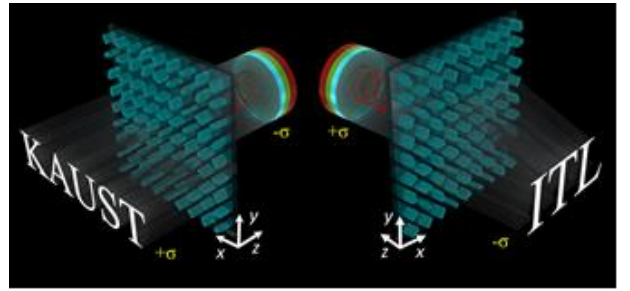

Fig. 1: Schematic diagram of proposed asymmetric wavefront shaping by utilizing single anisotropic unit element based metasurface. Proposed Metasurface efficiently reconstructs the dual holographic information of "KAUST" and "ITL" for orthogonal circular polarized light. Metasurface produces holographic image of "KAUST" when a "LCP" light is incident from backward direction. Similarly, when an "RCP" is incident from backward direction proposed metasurface produces holographic image of "ITL".

In addition to efficiency challenges, distinct information multiplexing on a single metasurface is a perilous task for the research community. Conventionally, segmented type metasurface design was adopted, e.g., Ansari *et al.* encoded two informations ("ITU" and "RHO") in single layer dielectric metasurface for orthogonal components of circular polarization [20]. Their designed surface achieved more than 50% efficiency for optical band of 549nm-700nm and maximum efficiency at 633nm. Another approach of multiple information encryption was utilized by Wan *et al.*, where they have encoded different holographic images for circular polarization by incorporating the idea of interleaved metasurfaces [21]. Such design strategies meet the information multiplexing requirements but suffer in terms of design space usage, i.e., half of the metasurface is utilized at a time and restrict the efficiency to 50%. Thus, dual-phase mergence (i.e., propagation and geometric phase) is proposed to improve the design space and enhance design efficiency. For example, Naveed *et al.* optimized an array of eight-unit elements to reconstruct on-axis asymmetric dual holograms for optical regime [22]. They have resolved the design space challenges but critically heightened the design complexities, which lead to fabrication constraints. Moreover, the measured efficiency for the operational wavelength of their device is 59%.

Here in this manuscript, we have proposed a highly efficient simplest single anisotropic unit element-based broadband dual information encryption technique for single-layer all-dielectric metasurfaces. A novel material (a-H:Si) with additional hydrogen content in conventional a-H:Si is utilized to have a wide transparent window in the optical regime with a high refractive index [23]. Moreover, only the geometric phase is exploited for full phase coverage (0-2π) and retains an additional degree of freedom (i.e., propagation phase). With the ability of multi-functionality with simplest design approach, proposed metasurface is capable of reproducing the holograms for wide range of optical regime with high efficiency compared to previously reported works [19-21]. The schematic illustration of the proposed concept is depicted in Fig. 1, where the metasurface reconstructs the two independent holographic images for orthogonal components of CP light under normal incidence. Metasurface produced a holographic image of "KAUST" when a left circularly polarized "LCP" light is incident from a backward direction. Similarly, when a right hand circularly "RCP" polarized light is incident from backward direction proposed metasurface produces a holographic image of "ITL" in the forward direction.

## II. Design Methodology

The subwavelength scale nano-resonators have a unique ability to manipulate the wavefronts and phase of incident light. Herein, to demonstrate the proposed helicity multiplexing concept, Panchartnam Berry (PB) phase is employed for covering the full phase range of 0 to 2π. For this purpose, a half waveplate is designed and optimized which can suppress the co-polarized light component and magnify the cross-polarized light component. Thus, we employed an anisotropic nano-resonating antenna of a-H:Si over glass substrate (SiO$_2$) as pictorially depicted in Fig. 2a.

Thus, for this purpose, the geometrical parameters of an anisotropic rectangular bar, i.e., length L, width W, height H, and period P, are carefully engineered. The optimal value of periodicity P=290 nm is designated using Nyquist criteria (i.e., P< (λ/2NA)) while keeping H= 400 nm constant for the sake of simplicity. L and W are tuned to achieve the functionality of the half-wave plate. The optimization graph of L and W for high cross-polarized transmissivity at three different optical wavelengths, i.e. 488 nm, 532 nm, and 633 nm, are depicted in Fig. 3 from (a) to (c) respectively. Their respective phase variations for cross-polarized transmission are depicted in Fig. 3 (d-f). By careful optimization, a single point with L = 215 nm and W = 80 nm was selected in such a way that it has simultaneously maximum cross-polarization transmission efficiency and minimum co-polarization transmission for all three operating wavelengths in order to maintain broadband functionality. Thus, when light impinges the unit element, the transmitted electric field ($E_t$) can be depicted by Equation (1):

$$E_t = R(\theta) \begin{pmatrix} e^{(i\varphi)} & 0 \\ 0 & e^{-(i\varphi)} \end{pmatrix} R^{-1}(\theta).E_i,$$
$$R(\theta) = \begin{pmatrix} Cos\theta & Sin\theta \\ Sin\theta & -Cos\theta \end{pmatrix} \quad (1)$$

where $E_i$ is incident electric field and and $R(\theta)$ is rational matrix multiplied with transformational matrix. Furthermore, "$\theta$" defines in-plane angular orientation of unit element as illustrated in Fig. 2b.

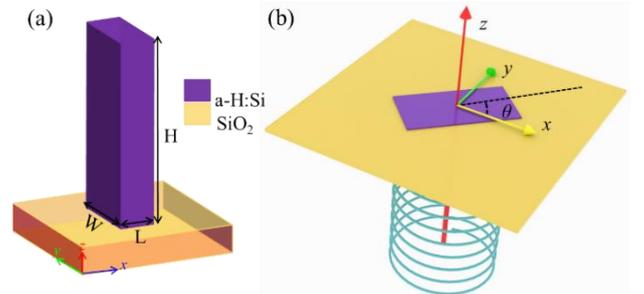

Fig. 2. Schematic view of optimized a-Si:H made nano-antenna that is used as a basic constituent element of metasurface. (a) 3D view of the nano-antenna. (b) A 2D perspective view of nano-antenna with the in-plane angular orientation defined by '$\theta$' with the *x*-axis.

After getting an optimized nano-antenna, the phase mask of two images, "ITL and KAUST," are calculated by using Gerchberg Saxton (GS) algorithm for 206 × 206 unit elements [24]. These phase masks are further processed by using Equation (2) to find a cumulative phase response that can be easily encrypted on the metasurface interface.

$$\Phi_t = \arg[\exp(i.\tan^{-1}\{\tan(\frac{\Phi_f - \Phi_b}{2})\})], \quad (2)$$

where $\Phi_t$ represents the total phase response, whereas $\Phi_f$ and $\Phi_b$ represents the phase of two images ("ITL" and "KAUST")

that we want to encode on the front and rear face of metasurface. Thus, the final phase imparted on metasurface using PB-phase is half of the total phase $\Phi_t$.

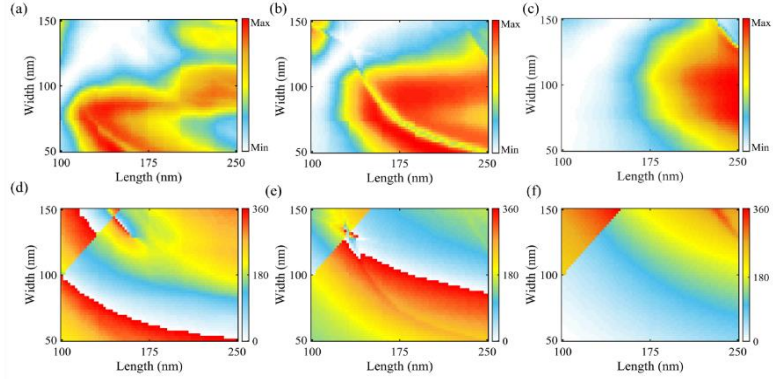

Fig. 3 Optimization of a-Si:H made nano-resonating antenna. (a-c) Cross-polarized transmission efficiency graphs as a function of nano-resonating antenna geometrical parameters at three different optical wavelengths of $\lambda_1 = 488$ nm, $\lambda_2 = 532$ nm, and $\lambda_3 = 633$ nm respectively. While (d-f) demonstrates the corresponding phase variations with respect to geometrical parameters. By simultaneous optimization of all three responses of transmission efficiency, a single optimal point of geometrical dimensions, i.e. L=215 nm and W=80 nm, is selected for metasurface design.

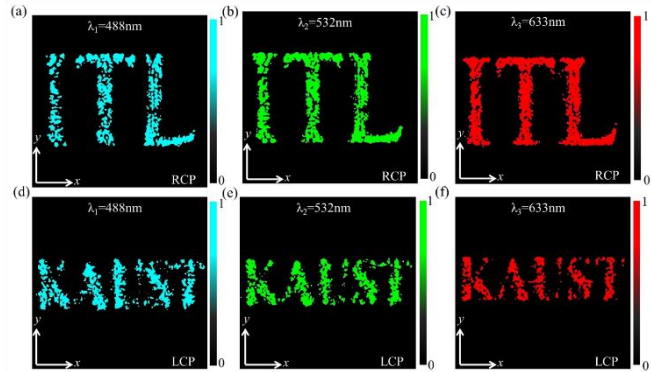

Fig. 4. Optical response of simulated metasurface. (a-c) demonstrate the highly efficient hologram reconstruction of logo "ITL" under RCP incidence for three different optical wavelengths, i.e., 488 nm, 532 nm, and 633 nm, respectively. While (d-f) represents a reconstruction of hologram "KAUST" under LCP incidence for all three operating wavelengths.

## III. RESULTS AND DISCUSSION

To validate the proposed theoretical spintronics-based design methodology for holography, we started with the numerical computation of phase masks of both pieces of information "KAUST" and "ITL" using a modified GS-algorithm for $60 \times 60$ μm$^2$ meta-hologram that corresponds to $206 \times 206$ array of nano-resonating elements. These calculated phase masks of images are then encoded in metasurface by employing PB-phase with a single anisotropic unit element of a-H:Si defined by Equation (1). All simulations are carried out in the commercially available tool Lumerical FDTD at three different optical wavelengths, i.e., 488 nm, 532 nm, and 633 nm. The simulated optical outputs of the reconstructed holograms of the designed metasurface are demonstrated in Fig. 4. The re-produced results quite well match the predicted one and verify the proposed methodology functionality for holographic meta-displays. Under LCP incidence, a holographic image "KAUST" is produced at a focal distance of 30 μm while the second holographic image of "ITL" is reconstructed for orthogonal component (RCP) incidence with high efficiency and fidelity. High efficiency with good fidelity and resolution of multiplexed hologram for wideband of optical regime proves the proposed strategy of the single element based spintronic based holography.

## IV. CONCLUSION

The phenomenon of spintronic-based asymmetric metaholograms is realized by exploiting low lose dielectric a-H:Si-based anisotropic nano-resonating antennas. The proposed simplest novel concept of asymmetric on-axis dual side hologram reconstruction breaks the constraints of design complexity and computation cost with fabrication ease. Benefiting from the PB-phase modulation of anisotropic nano unit element, the efficiency of design space is improved in terms of the degree of freedom and multi-functionality. Moreover, we have realized highly transmissive dual holograms in an optical regime based on the helicity and incidence direction of electromagnetic wave with simulated efficiencies of 55%, 75%, and 80% for the blue ($\lambda = 488$ nm), green ($\lambda = 532$ nm), and red light ($\lambda = 633$ nm) respectively. Thus, the proposed strategy of broadband spin multiplexed gives new direction toward many interesting applications of efficient displays, multi-media devices, imaging, AR/VR-based holographic elements, and data encryption.